\documentclass[aps,reprint,twocolumnsprl,showpacs,superscriptaddress,floatfix,notitlepage,nofootinbib]{revtex4-1}
\usepackage{amssymb,amsmath,amsfonts} 
\usepackage{epsfig,graphicx,physics}
\usepackage[colorlinks=true,citecolor=blue,linkcolor=red,pdftitle={Titulito},pdfauthor={Autores}]{hyperref}
\usepackage{mathtools}

\begin{document}
\title{Quantum thermalization mechanism and the emergence of symmetry-breaking phases}

\author{Sebastián Gómez}       \affiliation{Departamento de Estructura de la Materia, F\'{i}sica T\'{e}rmica y Electr\'{o}nica, Universidad Complutense de Madrid, Avenida Complutense s/n, E-28040 Madrid, Spain}
\affiliation{Instituto de Estructura de la Materia, IEM-CSIC, Serrano 123, E-28006 Madrid, Spain}

\author{\'{A}ngel L. Corps}
    \email[]{corps.angel.l@gmail.com}
    \affiliation{Institute of Particle and Nuclear Physics, Faculty of Mathematics and Physics, Charles University, V Hole\v{s}ovi\v{c}k\'{a}ch 2, 180 00 Prague, Czech Republic}
    
\author{Armando Rela\~{n}o}
    \email[]{armando.relano@fis.ucm.es}

    \affiliation{Departamento de Estructura de la Materia, F\'{i}sica T\'{e}rmica y Electr\'{o}nica, Universidad Complutense de Madrid, Avenida Complutense s/n, E-28040 Madrid, Spain}
    \affiliation{Grupo Interdisciplinar de Sistemas Complejos (GISC),
Universidad Complutense de Madrid, Avenida Complutense s/n, E-28040 Madrid, Spain}

\begin{abstract} 
We propose a generalization of the eigenstate thermalization hypothesis accounting for the emergence of symmetry-breaking phases. It consists of two conditions that any system with a degenerate spectrum must fulfill in order to thermalize. The failure of each of them generates a different non-thermalizing scenario. One is due to the absence of chaos and may indicate that extra constants of motion are required to describe equilibrium states. The other one implies the existence of initial conditions evolving towards symmetry-breaking equilibrium states. If it spreads across an entire spectral region, then this region gives rise to a symmetry-breaking phase. We explore the applicability of this formalism by means of numerical experiments on a three-site Bose-Hubbard model with two non-commuting discrete symmetries.
\end{abstract}

\maketitle

{\em Introduction.-} Understanding how an isolated quantum system thermalizes has been a fundamental goal since von Neumann's times \cite{Neumann1929}. Today, we have a solid theory for its underlying mechanism. The eigenstate thermalization hypothesis (ETH) states that every Hamiltonian eigenstate is thermal in chaotic quantum systems \cite{Deutsch1991,Srednicki1994,Rigol2008,Polkovnikov2011,Gogolin2016,DAlessio2016}. In its strongest version, the eigenstate expectation value of any few-body physical observable, $O_{nn} = \bra{E_n} \hat{O} \ket{E_n}$, agrees with the microcanonical ensemble in the thermodynamic limit (TL) \cite{Reimann2015,Yoshizawa2018,Sugimoto2021,Sugimoto2022},
\begin{equation}
  \label{eq:strongETH}
\textrm{max}_n \Delta_n(E) \equiv \textrm{max}_n \frac{|O_{nn}-O_{\textrm{ME}}(E)|}{\eta_O} \overset{N \rightarrow \infty}{\rightarrow}  0.
\end{equation}
Here, $O_{\textrm{ME}}(E)$ is the microcanonical average, and $\eta_O$ is the spectral width of the observable $\hat{O}$.

The reason why this guarantees thermalization lies in
the time-dependent expectation values for physical observables, $\langle \hat{O}(t) \rangle = \sum_{n,m} c_m^* c_n \, \textrm{e}^{-i(E_n-E_m)t/\hbar} \, O_{nm}$, from a pure initial condition, $\ket{\Psi(0)} = \sum_n c_n \ket{E_n}$. Assuming no degeneracies, off-diagonal terms in this expression dephase in the long-time limit, and hence the system approaches an effective equilibrium state, $\hat{\rho}_{\textrm{eq}}=\sum_n |c_n|^2 \ket{E_n} \bra{E_n}$, after a transient regime. If Eq. \eqref{eq:strongETH} holds and the initial condition is sufficiently narrow in energy, expectation values of physical observables in this effective equilibrium state coincide with microcanonical averages. There are strong arguments supporting that deviations from this equilibrium state are rare and short-lived
\cite{Reimann2012,Reimann2008,Linden2009,Linden2010,Reimann2010,Short2011,Short2012}.

Despite the universality of this mechanism \cite{Trotzky2012,Kaufman2016,Clos2016,Tang2018,Kota2011,Rigol2012,Ikeda2013,Beugeling2014,Steinigeweg2014,Kim2014,Jansen2019,Brenes2020,Richter2020,Wang2022}, whose exceptions are mainly due to integrability \cite{Kinoshita2006,Gring2012,Langen2015,Rigol2009,Essler2016,Mierzejewski2020}, many-body localization \cite{Choi2016,Smith2016,Oganesyan2007,Nandkishore2015,Corps2020,Corps2021}, the presence of scarred states \cite{Turner2018,Turner2018b,Schecter2019,Serbyn2021,Austin-Harris2025}, and certain other effects \cite{Relano2010,Shiraishi2017}, the scenario behind the non-degeneration hypothesis is highly involved. The level spacing in realistic quantum systems decreases exponentially with the number of particles, and thus dephasing times are expected to exceed the age of the Universe even for modest system sizes. In Srednicki's ansatz for the ETH \cite{Srednicki1994,Srednicki1999}, based on the random matrix theory (RMT) \cite{Mehta2004}, this problem is ruled out by assuming that off-diagonal matrix elements of physical observables also decay exponentially with the number of particles, $O_{nm} \propto \textrm{e}^{-S(E)/2}$, for $m \neq n$, where $S(E)$ is the thermodynamic entropy. But, despite an abundance of numerical evidence supporting this assumption \cite{Jansen2019,Richter2020,Wang2022,Beugeling2015,Mondiani2017,Nation2018,LeBlond2020}, it cannot be satisfied in some important physical situations, like symmetry-breaking phases. These are characterized by non-zero equilibrium values for order parameters, whose diagonal expectation values are necessarily zero \cite{Beekman2019}. This implies that some off-diagonal elements must contribute to the final equilibrium state, like the {\em tower of states} becoming degenerate in the TL when a continuous symmetry is broken \cite{Anderson1952,Bernu1992,Azaria1993,Bernu1994,Koma1994,Tasaki2019}. To complicate things further, the number of these states is typically very small \cite{note1}, and their role becomes relevant only under very specific circumstances ---for the right values of the system parameters and below the energy corresponding to the critical temperature of the transition. Thus, a number of fundamental questions emerge: What are the physical properties of the relevant degeneracies? Is there a simple way to predict in what spectral regions symmetry-breaking equilibrium states may appear? And considering that the distinction of diagonal and off-diagonal matrix elements is ill-defined in the presence of degeneracies, what can be said about the thermalization mechanism in this case?

In this Letter we provide answers to these questions. Our main result is a generalization of the ETH which accounts for the role of degeneracies and has the following properties: (i) it is basis-independent; (ii) it reduces to the standard ETH in the absence of degeneracies, and (iii) it distinguishes between two qualitatively different situations in which thermalization fails: the absence of chaos and symmetry breaking.

{\em Generalized ETH.-} We focus on isolated quantum systems with one or more symmetries. We label their eigenstates by two quantum numbers, $\ket{E_{n,\alpha}}$, where $\alpha$ accounts for the eigenstates of a (sub)set of commuting symmetries. We assume that our system is large enough so that $E_{n,\alpha} = E_{n,\beta} \equiv E_n$, $\forall \alpha, \beta$, for all practical purposes. We rule out remaining degeneracies by assuming that off-diagonal matrix elements between them fulfill Srednicki's ansatz.

Within this framework, let us consider a physical observable, $\hat{O}$, and the subspaces $\mathcal{H}_n$ spanned by Hamiltonian eigenstates with the same energy $E_n$, $\mathcal{H}_{n} \equiv  \textrm{span} \left\lbrace \ket{E_{n,\alpha}}:\/ \hat{H} \ket{E_{n,\alpha}} = E_n \ket{E_{n,\alpha}} \right\rbrace$, with $\alpha=1, \ldots, d_n$, where $d_n=\textrm{dim}\,\mathcal{H}_n$. We denote the eigenvalues of $\hat{O}$ within these subspaces as $\lambda^{(O)}_{n,\alpha}$, and its trace as $T^{(O)}_n$. We focus on an energy shell, $\mathcal{H}_{E,\delta E}$, centered at energy $E$ with subextensive width $2 \delta E$, defined as $\mathcal{H}_{E,\delta E} \equiv \textrm{span} \left\lbrace \ket{E_{n,\alpha}}:\/ |E_n - E| \leq \delta E \right\rbrace$. Then, we have:

Under the conditions above, every initial condition within $\mathcal{H}_{E,\delta E}$ thermalizes if and only if
 \begin{eqnarray}
   \label{eq:strong1}
   \textrm{max}_n \Delta_n^T(E) \equiv \textrm{max}_{n} \frac{\left| \frac{T^{(O)}_n}{d_n} - O_{\textrm{ME}}(E) \right|}{\eta_O} &\overset{N \rightarrow \infty}{\rightarrow}&  0, \\
   \label{eq:strong2}
\textrm{max}_{n,\alpha} \Delta_{n,\alpha}^{\lambda} (E) \equiv \textrm{max}_{n,\alpha} \frac{\left| \lambda^{(O)}_{n,\alpha} - \frac{T^{(O)}_n}{d_n} \right|}{\eta_O}  &\overset{N \rightarrow \infty}{\rightarrow}&  0.
\end{eqnarray}
 The proof is given in Appendix \ref{sec:proof}. 
 
 These equations reduce to Eq. \eqref{eq:strongETH} in the absence of degeneracies because $d_n=1$, $T^{(O)}_n=O_{nn}$, and Eq. \eqref{eq:strong2} is identically zero. Their physical relevance lies upon the following two scenarios:
 
 S1.- Failure of Eq. \eqref{eq:strong1}, which is due to non-chaotic behavior. In fully chaotic systems, Eq. \eqref{eq:strongETH} is expected to hold independently in each symmetry sector \cite{DAlessio2016,Santos2010a,Santos2010b,Gomez2011,Gubin2012}, and thus, Eq. \eqref{eq:strong1} is immediately fulfilled. Therefore, the failure of this equation implies that RMT does not work, so the system is not fully chaotic \cite{Bohigas1984}. This may be linked to the presence of additional constants of motion, implying that a generalized Gibbs ensemble (GGE) \cite{Jaynes1957,Jaynes1957b,Rigol2006} is required to describe equilibrium states.

 S2.- Failure of Eq. \eqref{eq:strong2}, which indicates symmetry-breaking when applied to order parameters, $\hat{M}$. As $T_n^{(M)}=0$ $\forall n$ for such operators \cite{Beekman2019}, this implies that some values of $\lambda_{n,\alpha}^{(M)}$ remain different from zero in the TL, allowing equilibrium states with $\langle \hat{M} \rangle \neq 0$. Therefore, we arrive at the following condition for symmetry-breaking phases:

 \textit{A spectral region is a symmetry-breaking phase if and only if it has no subspaces fulfilling Eq. \eqref{eq:strong2}}. 

 Under such circumstances, the application of the symmetry-generating operator to the eigenstate corresponding to $\lambda_{n,\alpha}^{(M)}$ necessarily results in the eigenstate corresponding to a different eigenvalue, $\lambda_{n,\beta}^{(M)}$, with $\beta \neq \alpha$ \cite{Beekman2019}. Thus, the different values of $\lambda_{n,\alpha}^{(M)}$ represent the different branches of the order parameter to which the system can equilibrate. And therefore, the condition formulated above implies that every initial condition equilibrates in one of these branches, or in a quantum superposition of them \cite{Corps2024}.
 
Besides this, Eq. \eqref{eq:strong2} necessarily fails when applied to the symmetry generators. If they are local operators, this indicates that they must be included in a GGE or a non-Abelian thermal state (NATS) \cite{Guryanova2016,Halpern2016,Halpern2020,Kranzl2023,Majidy2023,Murthy2023,Lasek2024,Patil2025} to describe symmetric equilibrium states.

{\em Physical insight.-} To illustrate our theory, we consider the three-site Bose-Hubbard model with periodic boundary conditions,
\begin{equation}
  \label{eq:model}
  \hat{H} = \sum_{i=1}^3  \left[ -J \left( \hat{a}^{\dagger}_{i+1} \hat{a}_i + \hat{a}_{i+1} \hat{a}^{\dagger}_i \right) + \frac{U}{N} \hat{a}^{\dagger}_i \hat{a}^{\dagger}_i \hat{a}_i \hat{a}_i \right],
\end{equation}
where $N$ is the number of particles, and $\hat{a}^{\dagger}_i$ and $\hat{a}_i$ are the creation and annihilation bosonic operators at site $i$. Eq. \eqref{eq:model} commutes with the total number of particles, $\hat{N}=\sum_{i}\hat{a}_{i}^{\dagger}\hat{a}_{i}$. 

The symmetries of this model are described by the dihedral group $D_3$, which has two non-commuting discrete generators: a $2\pi/3$ rotation, $\hat{R} \ket{n_1,n_2,n_3} = \ket{n_3,n_1,n_2}$, where $n_i$ indicates the number of particles in the site $i$; and a reflection, $\hat{S} \ket{n_1,n_2,n_3} = \ket{n_3, n_2, n_1}$ (see Appendix \ref{sec:classical}). $\hat{R}$ has three eigenvalues, $r\in \left\lbrace 1, \textrm{e}^{2 \pi i/3}, \textrm{e}^{4 \pi i/3} \right\rbrace$, and $\hat{S}$ has just two, $s\in\left\lbrace 1, -1 \right\rbrace$. This model also has a classical limit that can be easily obtained by taking $\hat{a}_{j} = \sqrt{N/2} \left( \hat{q}_{j} + i \hat{p}_{j} \right)$, and $N\to\infty$. The result is a classical Hamiltonian with three coordinates, $q_i$, and momenta, $p_i$, but just two degrees of freedom due to the conservation of particle number implying $\sum_i (q_i^2 + p_i^2)=2$.  In this way, $1/N$ plays the role of an effective Planck constant, $\hbar_{\textrm{eff}}\propto 1/N$, and therefore the TL can be studied by means of the classical equations. The physics of this model is quite complex, giving rise to persistent currents and vortex formation \cite{Tsubota2000,Scherer2007}, fragmented condensation \cite{Gallemi2015}, and a transition from regularity to chaos \cite{Garcia-March2018, delaCruz2020,Wozniak2022,Nakerst2023}; it is also a minimal model for a superfluid circuit \cite{Arwas2014}.

To explore the physical insight of our generalized ETH, we focus on two order parameters. First, the persistent current $\hat{C} = i \sum_j \left( \hat{a}^{\dagger}_{j+1} \hat{a}_j - \hat{a}^{\dagger}_j \hat{a}_{j+1} \right)$. As $\langle \hat{C} \rangle=0$ in eigenstates of $\hat{S}$, it is a good order parameter for an ordered phase in which the reflection symmetry is broken.

And second, the generalized imbalance $\hat{I} = \hat{n}_1 + \textrm{e}^{2 \pi i / 3} \hat{n}_2 + \textrm{e}^{4 \pi i / 3} \hat{n}_3$, which is such that $\langle \hat{I} \rangle / N=1$, $\textrm{e}^{2 \pi i/3}$ or $\textrm{e}^{4 \pi i/3}$ if all the particles are located at the first, the second or the third site. As $\langle \hat{I} \rangle=0$ in the eigenstates of $\hat{R}$, it is a good order parameter for an ordered phase in which the $2\pi/3$ rotation symmetry is broken. 

To study Eq. \eqref{eq:strong1}, we rely on the hopping between the first two sites, $\hat{h}_{12} = \hat{a}^{\dagger}_1 \hat{a}_2 + \hat{a}^{\dagger}_2 \hat{a}_1$.

Throughout the rest of this Letter, we always refer to the intensive versions of these three operators, $\hat{C}/N$, $\hat{I}/N$ and $\hat{h}_{12}/N$, whose spectral widths do not depend on system size.

The first step to explore the consequences of Eqs. \eqref{eq:strong1} and \eqref{eq:strong2} is to build the subspaces $\mathcal{H}_n$. We consider that $\alpha=\lbrace 0, 2 \pi/3, 4 \pi/3 \rbrace$ identifies the eigenvalues of $\hat{R}$, $r=\textrm{e}^{i \alpha}$; therefore $d_n=3$, $\forall n$. In the subspaces with $r=\textrm{e}^{2 \pi i/3}$ and $r=\textrm{e}^{4 \pi i/3}$, $\hat{S}$ and $\hat{R}$ do not commute. As a consequence, $E_{n, 2 \pi/3} = E_{n, 4 \pi/3}$, $\forall n$, so we can identify the breaking of both rotation and reflection with this choice. In the subspace with $r=1$, $\hat{R}$ and $\hat{S}$ are commuting. 

\begin{figure}[h!]
\includegraphics[width=0.53\textwidth]{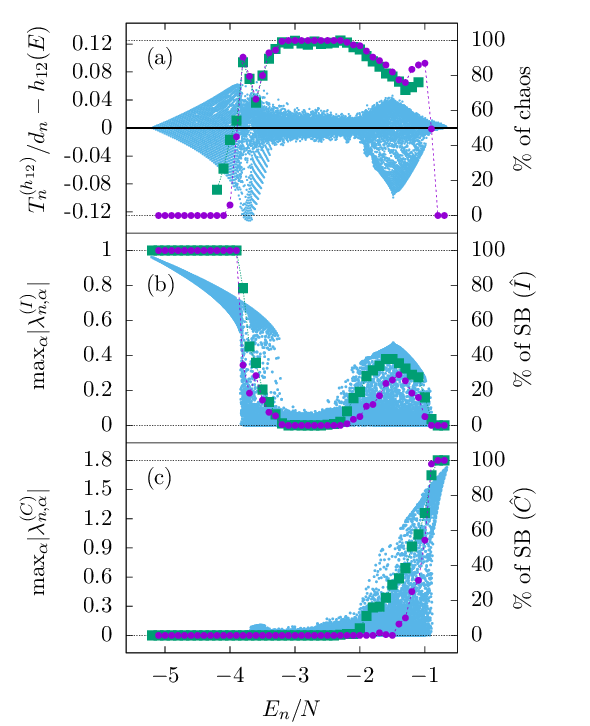}
    \caption{All the results are obtained with $N=320$, except where indicated. (a) Difference between the trace of $\hat{h}_{12}$ and its classical microcanonical value, and percentage of chaotic quantum and classical trajectories (purple and green points). (b) Maximum value of $|\lambda^{(I)}|$ (blue points), percentage of quantum and classical trajectories in which $\hat{I}$ is broken (purple points), and percentage of symmetry-breaking eigenvalues (green points). (c) Same as panel (b), for the current $\hat{C}$. The spectral statistics is performed with a system size of $N=800$.}
    \label{fig:eth}
\end{figure}

\begin{figure}[h!]
\hspace{-1cm}\includegraphics[width=0.53\textwidth]{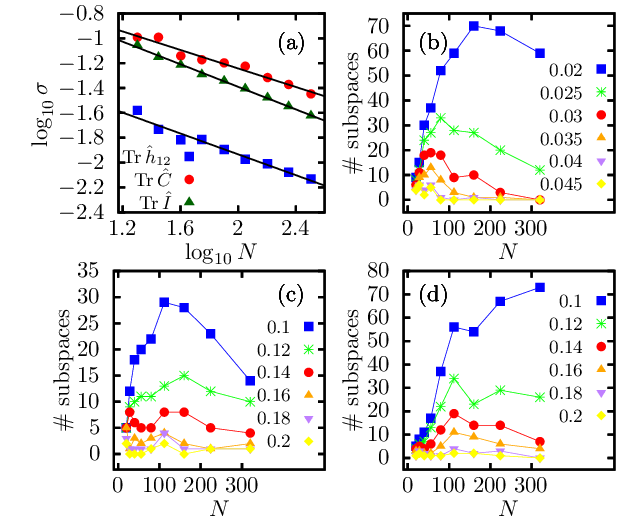}
    \caption{(a) Variance of the difference between the numerical (quantum) and microcanonical (classical) values for subspaces in the energy range $[-3.1,-2.3]$ for different observables. We find the following power-law behaviors: $\sigma\propto N^{-0.419}$ for $T_{n}^{(h_{12})}$ (squares), $\sigma\propto N^{-0.372}$ for $\hat{C}$ (circles) and $\sigma\propto N^{-0.454}$ for $\hat{I}$ (triangles). (b)-(d) Number of subspaces such that the difference between exact and microcanonical averages is below a given arbitrary bound (see legend), for $T_{n}^{(h_{12})}$ (b), $\hat{I}$ (c) and $\hat{C}$ (d). }
    \label{fig:scaling}
\end{figure}

For our numerical experiments, we set $U=-5$ and $J=1$. Our main result consists in identifying the following spectral regions:

{\em Thermalizing region}, with $-3.1 \lesssim E/N \lesssim -2.3$. First, spectral statistics and classical trajectories (see Appendix \ref{sec:detailsnumerical}) indicate that there is full chaos. According to S1, this implies that Eq. \eqref{eq:strong1} is necessarily fulfilled. In Fig. \ref{fig:eth}(a) we show that the trace of $\hat{h}_{12}$ always fluctuates around the microcanonical average, obtained from the classical version of Eq. \eqref{eq:model} (see Appendix \ref{sec:classical} for details). A deeper test is shown in Fig. \ref{fig:scaling}. In panel (a), we can see that the width of the distribution of $T_n^{(\hat{h}_{12})}/d_{n} - h_{12}(E)$ decreases as a power law, $\sigma \propto N^{-0.419}$. This decreasing behavior is close to Srednicki's ansatz, in which diagonal expectation values fluctuate around the microcanonical with a width proportional to $e^{-S(E)/2}=[\rho(E)]^{-1/2} \propto N^{-1/2}$. In Fig. \ref{fig:scaling}(b) we study the total number of subspaces in which $|T_n^{(\hat{h}_{12}}/d_{n} - h_{12}(E)|$ is larger than several bounds. Eq. \eqref{eq:strong1} requires that this number goes to zero in the TL. Results in Fig. \ref{fig:scaling}(b) suggest the existence of a general pattern: first, this number increases with $N$, then it reaches a maximum, and finally it decreases for larger system sizes; the smaller the bound, the larger the number of particles required to reach the maximum. Although much larger systems would be needed to generalize this result to much lower bounds, the overall picture suggests that Eq. \eqref{eq:strong1} is fulfilled.

As RMT makes no statements about the correlations between different symmetry sectors, full chaos does not imply Eq. \eqref{eq:strong2}. Nevertheless, results in Figs. \ref{fig:eth}(b) and \ref{fig:eth}(c) are compatible with $\lambda_{n,\alpha}$ fluctuating around zero for both order parameters. We can see in Fig. \ref{fig:scaling}(a) that the width of the corresponding distributions decreases following a power law in both cases. And Figs. \ref{fig:scaling}(c) and \ref{fig:scaling}(d) suggest a similar pattern for $\lambda^{(I)}_{n,\alpha}$ and $\lambda^{(C)}_{n,\alpha}$ that for the trace of $\hat{h}_{12}$, although the results for $\hat{C}$ are less conclusive because of the behavior of the lower bound (diagonalizing larger systems would be necessary to get a clearer image).

All these results suggest that both Eqs. \eqref{eq:strong1} and \eqref{eq:strong2} are fulfilled within this region. Thus, the standard microcanonical ensemble is expected to hold, and therefore no symmetry breaking can be observed.

\begin{figure}[h!]
\includegraphics[width=0.53\textwidth]{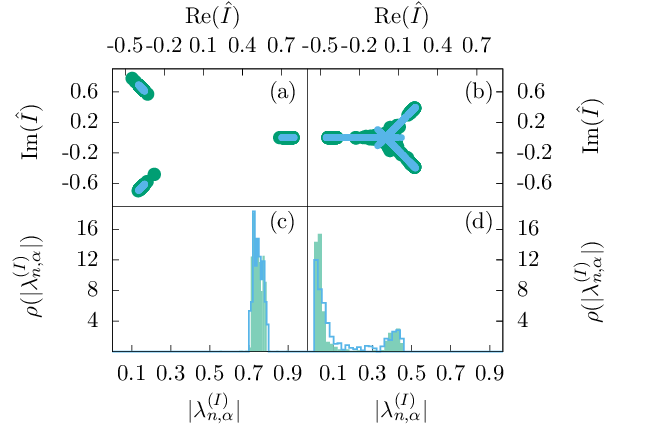}
    \caption{(a)-(b) Complex eigenvalues of $\hat{I}$ (blue) and long-time average of different classical trajectories (green) for $E/N=-4.2$ (a) and $E/N=-1.4$ (b). Classical results correspond to $1000$ trajectories in an interval of width $0.1$ around $E/N$. (c)-(d) Distribution of absolute value of the eigenvalues of $\hat{I}$ (blue) and classical average (green). Quantum results correspond to $N=320$.}
    \label{fig:imbalance}
\end{figure}

{\em Rotation-breaking region}, with $E/N \lesssim -3.9$. Fig. \ref{fig:eth}(b) shows that none of the subspaces fulfill Eq. \eqref{eq:strong2} for $\hat{I}$. From S2, we conclude that this is an ordered phase in which the rotational symmetry is broken. The analysis of the classical trajectories confirms this statement. All the initial conditions remain trapped in different (and apparently disconnected) regions of the phase space, giving rise to non-zero values for the long-time average of $I(\mathbf{q},\mathbf{p})/N$. Fig. \ref{fig:imbalance}(a) shows that the values of $\lambda^{(I)}_{n,\alpha}$ at $E/N \sim -4.2$ are distributed in three different regions linked by $2\pi/3$ rotations (the same qualitative results are obtained at other energies). We can also see that all the classical trajectories equilibrate in one of these regions. Furthermore, Fig. \ref{fig:imbalance}(c) shows that the distributions for the classical and the quantum results are very similar. Therefore, we conclude that the three different branches of the order parameter consist of one of the sites being more populated than the other, equally populated, two. 

Finally, results in Fig. \ref{fig:eth}(a) and \ref{fig:eth}(b) show that there is no clear link between the failure of Eq. \eqref{eq:strong2} and the absence of chaos. The degree of both classical and quantum chaos changes within this region, from apparent integrability in the lowest energy region, to around $50 \%$ of chaos at $E/N \sim -3.9$. (We do not show the result for spectral statistics below $E/N \sim -4.2$ because the level density is too low to obtain significant values with steps of $\Delta E/N=0.1$; a global fit to $E/N<-4.2$ results in $0 \%$ of quantum chaos).

{\em Reflection-breaking region}, with $E/N \gtrsim -0.9$. Fig. \ref{fig:eth}(c) shows that none of the subspaces fulfill Eq. \eqref{eq:strong2} for $\hat{C}$,
 and all the classical trajectories relax to equilibrium values with a non-zero value for $C$. Thus, this is a small reflection-breaking region, characterized by the presence of persistent currents. At first sight, Fig. \ref{fig:eth}(a) suggests that there is no chaos within this region. Nevertheless, it is too small to get significant results for the quantum version of the system, and therefore to reach a safe conclusion about its degree of chaos.

{\em Mixed regions}, with $-3.8 \lesssim E/N \lesssim -3.1$ and $-2.3 \lesssim E/N \lesssim -0.9$. In Fig. \ref{fig:eth}(b) we can see that $\lambda^{(I)}_{n,\alpha}$ fluctuates around zero in some subspaces, whereas it is significantly different from zero in some others. A similar behavior is observed for $\lambda^{(C)}_{n,\alpha}$ in Fig. \ref{fig:eth}(c) for $-2.3 \lesssim E/N \lesssim -0.9$; on the contrary, there is no reflection breaking below $E/N \sim -2.3$. The results for the classical trajectories agree with this fact: we can find initial conditions relaxing to either symmetric and symmetry-breaking equilibrium states. This conclusion is reinforced by Figs. \ref{fig:imbalance}(b) and \ref{fig:imbalance}(d), which show results for $E/N \sim -1.4$. Again, the non-zero values of $\lambda^{(I)}_{n,\alpha}$ are distributed over three different regions linked by $2\pi/3$ rotations. And there is again a very good agreement between the classical equilibrium values and the possible values of $\lambda^{(I)}_{n,\alpha}$. In particular, Fig. \ref{fig:imbalance}(d) suggests that, depending on the initial condition, we can reach a symmetric equilibrium value, represented by the large peak close to zero for both $\lambda^{(I)}_{n,\alpha}$ and the classical long-time average, or a rotation-breaking value, characterized by the modulus of $\lambda^{(I)}_{n,\alpha}$ and the classical long-time average being around $0.4$.
Contrary to what happens in the fully rotation-breaking region, here the symmetry-breaking equilibrium states have one site less populated than the other, equally populated, two.

The behavior of $T^{(h_{12})}/d_n$, shown in Fig. \ref{fig:eth}(a), is a bit more complicated. Some subspaces show an ordered pattern, typical of integrable systems, whereas some others show erratic and narrow fluctuations, typical of chaos \cite{Peres1984}. This is compatible with the degree of chaos found in both the classical and the quantum versions of the system. However, the microcanonical average does not coincide with the erratic part of $T^{(h_{12})}/d_n$. This is probably because the microcanonical average is obtained by integrating all the available phase space, including both the chaotic and the non-chaotic regions.

\textit{Discussion.-} The main contribution of this Letter is a generalization of the ETH which gives rise to two non-thermalizing scenarios. The second one implies the existence of symmetry-breaking equilibrium states. By means of a numerical experiment on a quantum system with a clear classical analogue, we have shown that such equilibrium states may appear in two different situations. First, they can completely characterize a spectral region, giving rise to a symmetry-breaking phase. Second, they may be surrounded by symmetric equilibrium states, thus playing a role similar to many-body scars within a chaotic sea. 

\textit{Acknowledgments.-} A. L. C. acknowledges fruitful discussions with P. Cejnar, P. Stránský and J. Novotný. A.R. acknowledges financial support by the Spanish grant
PID2022-136285NB-C31 funded by Ministerio de
Ciencia e Innovación/Agencia Estatal de Investigación MCIN/AEI/10.13039/501100011033 and FEDER “A Way of Making Europe”. A.L.C. acknowledges financial support from the Czech Science Foundation under project No. 25-16056S as well as the JUNIOR UK Fund project carried out at the Faculty of Mathematics and Physics, Charles University.

\appendix

\section{Proof of the main result}\label{sec:proof}

Here we prove that Eqs. \eqref{eq:strong1} and \eqref{eq:strong2} hold if and only if all the initial conditions narrow enough in energy thermalize.

{\em Forward implication.-} Let us consider that the system starts in an initial condition, $\ket{\Psi(0)}=\sum_{n,\alpha} c_{n,\alpha} \ket{E_{n,\alpha}}$, where $c_{n,\alpha} \neq 0$ only in a subextensive interval $[E-\delta E, E+\delta E]$, and let us calculate the resulting long-time average of a physical observable $\hat{O}$. To do so, we consider the matrix for $\hat{O}$ in a subspace of energy $E_m$, $\hat{O}^{(m)}_{\alpha \beta} = \bra{E_{m,\alpha}} \hat{O} \ket{E_{m,\beta}}$, with $\alpha,\beta \in \{1, \ldots, d_m\}$. Then, we diagonalize it as $\hat{O}^{(m)} = \hat{P}^{(m)} \hat{D}^{(m)} (\hat{P}^{(m)})^{-1}$, where $\hat{P}$ is a block-diagonal matrix that acts as the identity in every subspace with energy different from $E_m$. The result of repeating this procedure on every subspace is a new eigenbasis for the Hamiltonian, $\{ \ket{\widetilde{E}_{m,\alpha}} \}$, in which $\hat{O}^{(m)}_{\alpha,\beta}$ is diagonal in every subspace. Then, writing the initial condition in this eigenbasis, $\ket{\Psi(0)} = \sum_{n,\alpha} \widetilde{c}_{n,\alpha} \ket{\widetilde{E}_{n,\alpha}}$ we get
\begin{equation}
\begin{split}
\bra{\Psi(t)} \hat{O} \ket{\Psi(t)} &= \sum_{n,m} \sum_{\alpha,\beta} \widetilde{c}^*_{m,\beta} \, \widetilde{c}_{n,\alpha} \, \textrm{e}^{-i(E_n - E_m) t/\hbar} \cdot \\ &\cdot \bra{\widetilde{E}_{m,\beta}} \hat{O} \ket{\widetilde{E}_{n,\alpha}}. 
\end{split}
\end{equation}
Thus, if we consider that $\bra{\widetilde{E}_{m,\beta}} \hat{O} \ket{\widetilde{E}_{n,\alpha}}$ decays exponentially with the system size for $n \neq m$, we can dismiss these off-diagonal terms in the calculation of the long-time average. Therefore:
\begin{equation}
\label{eq:long-time}
\begin{split}
\overline{\hat{O}} &= \sum_n \sum_{\alpha,\beta} \widetilde{c}^*_{n,\beta} \, \widetilde{c}_{n,\alpha} \bra{\widetilde{E}_{n,\beta}} \hat{O} \ket{\widetilde{E}_{n,\alpha}}= \\ &= \sum_{n,\alpha} |\widetilde{c}_{n,\alpha}|^2 \lambda_{n,\alpha},
    \end{split}
\end{equation}
where $\lambda_{m,\alpha}, \, \alpha \in \{1, \ldots, d_m\}$ are the eigenvalues of $\hat{O}^{(m)}_{\alpha,\beta}$. 

Finally, if both Eqs. \eqref{eq:strong1} and \eqref{eq:strong2} are satisfied, then
\begin{equation}
\textrm{max}_{n,\alpha} \frac{| \lambda^{(O)}_{n,\alpha} - O(E) |}{\eta_{\hat{O}}} \rightarrow 0, \; N \rightarrow \infty.
\end{equation}
And therefore, the long-time average given in Eq. \eqref{eq:long-time} coincides with the microcanonical average.

{\em Backward implication.-} This is equivalent to showing that if (a) $\max_{n}|T_{n}^{(O)}/d_{n}-O_{\textrm{ME}}(E)|\nrightarrow 0$ or (b) $\max_{n,\alpha}|\lambda_{n,\alpha}^{(O)}-T_{n}^{(O)}/d_{n}|\nrightarrow 0$ when $N\to\infty$, then there exists at least one non-thermalizing initial condition. 

Let us begin by assuming that (a) is true. This means that there exists at least one subspace with energy $E_m$ in which
\begin{equation}\label{eq:(2)}
    \left|\frac{T_{m}^{(O)}}{d_{m}}-O_{\textrm{ME}}(E)\right|
    \nrightarrow 0, \; N \rightarrow \infty
\end{equation}
Thus, it is enough to find an initial condition $\ket{\Psi(0)}$ such that $\overline{\hat{O}}=T_{m}/d_{m}$ to show that condition (a) implies the existence of at least one non-thermalizing initial condition. We can build a simple one by means of $\ket{\Psi(0)} = \sum_{\alpha=1}^{d_m} \ket{\widetilde{E}_{m,\alpha}}/\sqrt{d_m}$. We have then that $|| \ket{\Psi(0)} ||=1$, and $\ket{\Psi(t)} = \sum_{\alpha=1}^{d_m} \textrm{e}^{-i E_m t/\hbar} \, \ket{\widetilde{E}_{m,\alpha}}/\sqrt{d_m}$. Then, 
\begin{equation}
\begin{split}
\overline{\hat{O}}&=\bra{\Psi(0)}\hat{O}\ket{\Psi(0)}=\sum_{\alpha,\beta=1}^{d_{m}}\frac{\bra{\widetilde{E}_{m,\beta}}\hat{O}\ket{\widetilde{E}_{m,\alpha}}}{d_{m}}\\&=\sum_{\alpha=1}^{d_{m}}\frac{\lambda_{m,\alpha}}{d_{m}}=\frac{T_{m}^{(O)}}{d_{m}},
\end{split}
\end{equation}
which proves (a). 

Let us now assume that (b) is satisfied. Similarly to case (a), we can assume that there exists an eigenspace of energy $E_{m}$ and an eigenvalue $\lambda_{n,m}^{(\hat{O})}$ such that
\begin{equation}
    \left|\lambda_{n,m}^{(O)}-\frac{T_{m}^{(\hat{O})}}{d_{m}}\right|\nrightarrow 0,\,\,\,N\to\infty.
\end{equation}
Due to our previous results, we may assume that (a) is not satisfied and that $\max_{m}|T_{m}^{(\hat{O})}/d_{m}-O_{\textrm{ME}}|\rightarrow 0$. Let $\lambda_{\alpha}^{(m)}$ be the eigenvalue of $\hat{O}^{(m)}$ such that $\bra{\widetilde{E}_{m,\beta}} \hat{O} \ket{\widetilde{E}_{m,\alpha}}=\lambda_{m,\alpha}$ and let us consider an initial condition $\ket{\Psi(0)}=\ket{\widetilde{E}_{m,\alpha}}$. Thus, the long-time average of the expectation value of $\hat{O}$ is simply 
\begin{equation}
\overline{\hat{O}}=\bra{\widetilde{E}_{m,\alpha}}\hat{O}\ket{E_{m,\alpha}}=\lambda_{m,\alpha}^{(O)}
\end{equation}
And this initial state does not thermalize.

\section{Some details of the model}\label{sec:classical}

{\em Classical limit.--} The three-site Bose-Hubbard model, Eq. \eqref{eq:model}, has a well-defined classical limit when $N\to\infty$. In this limit, the bosonic positions and momenta commute, $[\hat{q}_{i},\hat{p}_{j}]\propto 1/N\to0$, so they can be treated as classical variables defined on the phase space, $(\mathbf{q},\mathbf{p})\in\mathcal{M}$. The simple substitution $\hat{a}_{j}\to \sqrt{N/2}(q_{j}+ip_{j})$ allows us to obtain a classical Hamiltonian from the quantum Hamiltonian, $\hat{H}\to H(\mathbf{q},\mathbf{p})$, 
\begin{equation}
    \frac{H}{N}=\sum_{i=1}^{3}\left[-J(q_{i+1}q_{i}+p_{i+1}p_{i})+\frac{U}{4}(q_{i}^{2}+p_{i}^{2})^{2}\right],
\end{equation}
where $\sum_{i}(q_{i}^{2}+p_{i}^{2})\leq 2$. One can similarly obtain classical expressions for the current $\hat{C}$, the imbalance $\hat{I}$, and the hopping $\hat{h}_{12}$, namely
\begin{equation}
    \frac{I}{N}=\frac{1}{2}\left[(q_{1}^{2}+p_{1}^{2})+e^{2\pi i/3}(q_{2}^{2}+p_{2}^{2})+e^{4\pi i/3}(q_{3}^{2}+p_{3}^{2})\right],
\end{equation}
\begin{equation}
    \frac{C}{N}=\sum_{j}(p_{j}q_{j+1}-p_{j+1}q_{j}),
\end{equation}
and
\begin{equation}
    \frac{h_{12}}{N}=q_{1}q_{2}+p_{1}p_{2}.
\end{equation}

\textit{Symmetries.--} Eq. \eqref{eq:model} is invariant under $2\pi/3$ and $4 \pi/3$ rotations, and under the reflection around any of its symmetry axes. This set of transformations constitutes the dihedral $D_3$ group, whose generators are the operators $\hat{R}$ and $\hat{S}$.

\section{Details of numerical calculations}\label{sec:detailsnumerical}

\textit{Evaluation of quantum chaos.--}
We rely on the distribution of nearest-neighbor level spacings, $P(s)$. In the case of semiclassical systems exhibiting the coexistence of chaotic and regular phase space regions, Berry and Robnik \cite{Berry1984} derived an expression for $P(s)$ assuming that eigenlevel sequences from chaotic regions follow the Wigner-Dyson distribution \cite{Mehta2004,Bohigas1984} while sequences from regular regions follow the Poisson distribution \cite{Berry1997},
\begin{equation}\label{eq:psberryrobnik}
    P(s;\rho)=\rho^{2}e^{-\rho s}\textrm{erfc}(\frac{1}{2}\sqrt{\pi}\overline{\rho}s)+(2\rho\overline{\rho}+\frac{1}{2}\pi\overline{\rho}^{3}s)e^{-\rho s-\frac{1}{4}\pi\overline{\rho}^{2}s^{2}},
\end{equation}
where $\rho$ is the fraction of Poissonian (regular) sequences and $\overline{\rho}=1-\rho$. Classically, $\rho$ represents the proportion of regular orbits. Taken as a one-parameter function, Eq. \eqref{eq:psberryrobnik} may be fitted to the numerical $P(s)$ to assess the degree of chaos through $\rho$.

As the universal predictions of the RMT only apply if $\langle s \rangle =1$, a prior mandatory step is the so-called unfolding procedure, which removes system-dependent features in the energy spectrum \cite{Gomez2002,Corps2020b}. Our method consists of a numerical fit to a polynomial of degree $4$ to the cumulative level density, which gives rise to a smooth function $\overline{N}(E)$. Then, the unfolded energy levels are obtained as $\epsilon_n = \overline{N}(E_n)$, from which we calculate $s_n = \epsilon_{n+1} - \epsilon_n$, which automatically fulfills $\langle s \rangle=1$.  In our numerical experiments, we have divided the spectrum of the Bose-Hubbard model into equiespaced sequences with width $\delta E=0.1$, performed the unfolding procedure to each of them separately, and calculated the fraction $\rho$ reported in Fig. \ref{fig:eth}(a). We have worked with $N=800$, $r=1$ and $s=\pm 1$. 

{\em Evaluation of classical chaos.--} For a given energy $E$, we randomly generate 200 initial conditions $(\mathbf{q},\mathbf{p})$ with a uniform distribution fulfilling $\sum_{i}q_{i}^{2}+p_{i}^{2}=2$; to do so we consider all initial conditions such that $H(\mathbf{q},\mathbf{p})/N\in[E-\delta E,E+\delta E]$ with $\delta E=0.05$. Then, we numerically solve the Hamilton equations for each initial condition with two different computational precisions (with $20$ and $25$ significant figures), up to time $t=1000$. Finally, we compute the time average of the distance of these two trajectories, for each initial condition. We consider the trajectory as chaotic if this average is greater than $0.3$. (Similar results are obtained with other bounds). This is in agreement with one of the most common definitions of classical chaos based on the exponential sensitivity of nearby initial conditions \cite{Gutzwillerbook}. The percentage of chaotic trajectories for a given energy is reported in Fig. \ref{fig:eth}(a). 

The classical microcanonical average at a given energy $E$ in Fig. \ref{fig:eth}(a) is 
\begin{equation}
    \langle h_{12}\rangle_{\textrm{ME}}=\frac{\int \textrm{d}\mathbf{q}\textrm{d}\mathbf{p}\,h_{12}(\mathbf{q},\mathbf{p})\delta[H(\mathbf{q},\mathbf{p})-E]}{\int \textrm{d}\mathbf{q}\textrm{d}\mathbf{p}\,\delta[H(\mathbf{q},\mathbf{p})-E]}
\end{equation}
where $(\mathbf{q},\mathbf{p})\in\mathcal{M}$. The integrals in this equation have been computed numerically through Montecarlo simulations, considering $10^{12}$ points in phase space, and an energy width $\delta E =0.01$.

\textit{Evaluation of symmetry-breaking.--} To compute the percentage of symmetry-breaking classical trajectories in Fig. \ref{fig:eth}(b)-(c), we consider the long-time average of each order parameter. We consider that a trajectory is symmetry-breaking if the absolute value of this magnitude is greater than its maximum possible value divided by $10$. Thus, this bound is $0.1$ for $\hat{I}$ and $\sqrt{3}/10$ for $\hat{C}$. 

\textit{Eigenvalues $\lambda_{n,\alpha}^{(O)}$ and traces $T_{n}^{(O)}$.--}  In Fig. \ref{fig:eth} and Fig. \ref{fig:imbalance} we report the values of the eigenvalues $\lambda_{n,\alpha}^{(\hat{O})}$ and the traces $T_{n}^{(O)}$ for $\hat{h}_{12}$, $\hat{I}$ and $\hat{C}$. To compute these quantities, first we diagonalize the Hamiltonian in each of the rotation subspaces. These subspaces are built considering the two eigenstates that are exactly degenerate (states belong to the $r=e^{2\pi i/3}$ and $r=e^{4\pi i/3}$ sectors, which do not commute with the reflection symmetry) and the eigenstate with $r=1$ such that its energy is closest to that of the previous states. Finally, we compute the matrices $\hat{h}_{12}$, $\hat{I}$ and $\hat{C}$ in these subspaces, and calculate their eigenvalues and traces numerically.


\begin{thebibliography}{100}
\bibitem{note1} The scenario becomes even more complicated when considering that the gap between the states belonging to the tower decays as $1/N$, being $N$ the number of particles, whereas the decay of the mean level spacing is exponential. In the case of discrete symmetries both decays are exponential.
\end{thebibliography}
\end{document}